\documentclass[11pt]{article}

\usepackage[letterpaper,margin=0.75in]{geometry}

\usepackage{microtype}

\usepackage{titlesec}
\titleformat{\section}
  {\normalfont\large\bfseries}
  {\thesection}
  {0.75em}
  {}

\titleformat{\subsection}
  {\normalfont\normalsize\bfseries}
  {\thesubsection}
  {0.75em}
  {}

\usepackage{authblk}

\usepackage{amsmath,amssymb,amsfonts}
\usepackage{graphicx}
\usepackage[Final]{changes}

\usepackage{enumitem}
\usepackage{array}

\usepackage{cuted}

\usepackage[numbers,sort&compress]{natbib}
\usepackage{hyperref}
\hypersetup{
    colorlinks=true,
    linkcolor=blue,
    citecolor=blue,
    urlcolor=blue
}

\title{Spatiotemporal Tracking of Optical Speckles in Turbulent Atmospheric Propagation}

\author[1]{Travis M. Crumpton}
\author[1,2,3,4,*]{Luat T. Vuong}

\affil[1]{Department of Electrical and Computer Engineering, University of California, Riverside, California 92521, USA}
\affil[2]{Department of Mechanical Engineering, University of California, Riverside, California 92521, USA}
\affil[3]{Materials Science and Engineering Program, University of California, Riverside, California 92521, USA}
\affil[4]{Department of Physics, University of California, Riverside, California 92521, USA}
\affil[*]{Corresponding author: luatv@ucr.edu}

\date{}

\begin{document}

\twocolumn[
\begin{@twocolumnfalse}

\maketitle

\begin{abstract}
The speckle fields produced by optical beam propagation through atmospheric turbulence are typically described using ensemble-averaged intensity and coherence statistics, which obscure the speckle-level dynamics. Here, we investigate the spatiotemporal evolution of speckles generated during Gaussian beam propagation through turbulence by explicitly tracking them as discrete resolvable substructures. We quantify object-level persistence, transverse extent, and trajectories as functions of propagation distance, turbulence strength, and source-plane beam width. Under fixed detection criteria, we observe that a subset of these speckles exhibits measurable width statistics and persistence distributions whose form depends on turbulence strength and source-plane beam size. This object-level framework remains well-defined within the regime of strong turbulence and provides a complementary approach to characterizing turbulent propagation beyond conventional ensemble-averaged metrics.
\end{abstract}

\begin{center}
\small
This manuscript is the author-prepared version of an article published in
\textit{Optica Applied Optics}. The final published version is available at:
\href{https://doi.org/10.1364/AO.599858}{https://doi.org/10.1364/AO.599858}.
\end{center}

\vspace{1em}

\end{@twocolumnfalse}
]

\section{Introduction}

Speckle fields arise when a coherent electromagnetic wave encounters random phase or amplitude modulations imposed by rough surfaces, inhomogeneous media, or optical turbulence in atmospheric and oceanic environments \cite{korotkova2002speckle, korotkova2002speckle2, long2023underwater, beason2019statistical}. Their statistical behavior has been extensively studied using first- and second-order ensemble descriptions, particularly through the foundational work of Goodman \cite{goodman1976some, goodman2007speckle} and Andrews \cite{andrews1999theory,andrews2005laser, andrews2023laser}. These approaches characterize speckle through intensity distributions, spatial correlation functions, and scintillation metrics. Extensions to spatiotemporal correlations \cite{liguo2023spatiotemporal,goldfischer1965autocorrelation}, phase-singularity structures \cite{angelsky2009spatial,cheng2014phase}, and propagation dynamics \cite{lee1976statistics} have further refined this statistical framework. 

In practical applications, speckle is commonly treated either as an artifact to be mitigated using adaptive optics or filtering techniques or as a statistical signal used to characterize propagation environments or optical fields. Speckle reduction strategies such as digital filtering \cite{ozcan2007speckle}, polychromatic speckle mitigation \cite{van2020improved}, and statistical post-processing \cite{gladysz2008detection} have previously been demonstrated to increase performance of an optical system in the presence of speckle patterns. Characterization methods in static samples include the analysis of speckle field phase singularity dynamics under frequency variation \cite{cheng2014phase, zhang2007speckle}, fractal analysis of spatial speckle intensity fluctuations \cite{guyot2004spatial}, and statistical analysis of binarized speckle patterns \cite{kayahan2010measurement}. In temporally evolving systems, speckle is studied via intensity covariance \cite{garnier2018imaging}, temporal contrast \cite{garcia2025experimental}, speckle decorrelation lifetimes \cite{macintosh2005speckle}, and statistical modeling of speckle intensity distributions \cite{migliaccio2019sar, migliaccio2024link}. In many cases, the emphasis remains on ensemble statistics rather than on the evolution of individual intensity features.

Despite this progress, global ensemble-averaged descriptions inherently suppress information about the evolution of individual speckle structures by collapsing local interference dynamics into global statistical observables. During turbulent propagation, cumulative phase distortions fragment the optical field into resolvable substructures that evolve within a globally diverging beam envelope. While temporal fluctuations of speckle fields are known to encode properties of the intervening medium, ensemble metrics do not directly resolve the temporal object-level dynamics of these substructures.

Here, we investigate the spatiotemporal evolution of individual speckles produced during turbulent propagation of a Gaussian beam. By explicitly tracking speckles as discrete objects, we quantify their transverse width, trajectories, and persistence (lifetime based on creation and disappearance events) as functions of propagation distance. We examine how the persistence and width statistics observed using this object-level framework evolve across turbulence strengths spanning weak to strong turbulence conditions \cite{andrews1999theory, andrews2005laser, andrews2023laser}. Our analysis focuses on the roles of the beam width $w_0$ and the turbulence strength $C_n^2$, while effects associated with differing inner scale $\ell_0$ and outer scale $L_0$ are not considered here. We complement traditional ensemble metrics with an object-level analysis framework. This study introduces a feature-resolved approach to examining speckle evolution in turbulent propagation.

\begin{figure*}[htbp]
  \centering
  \includegraphics[width=\linewidth]{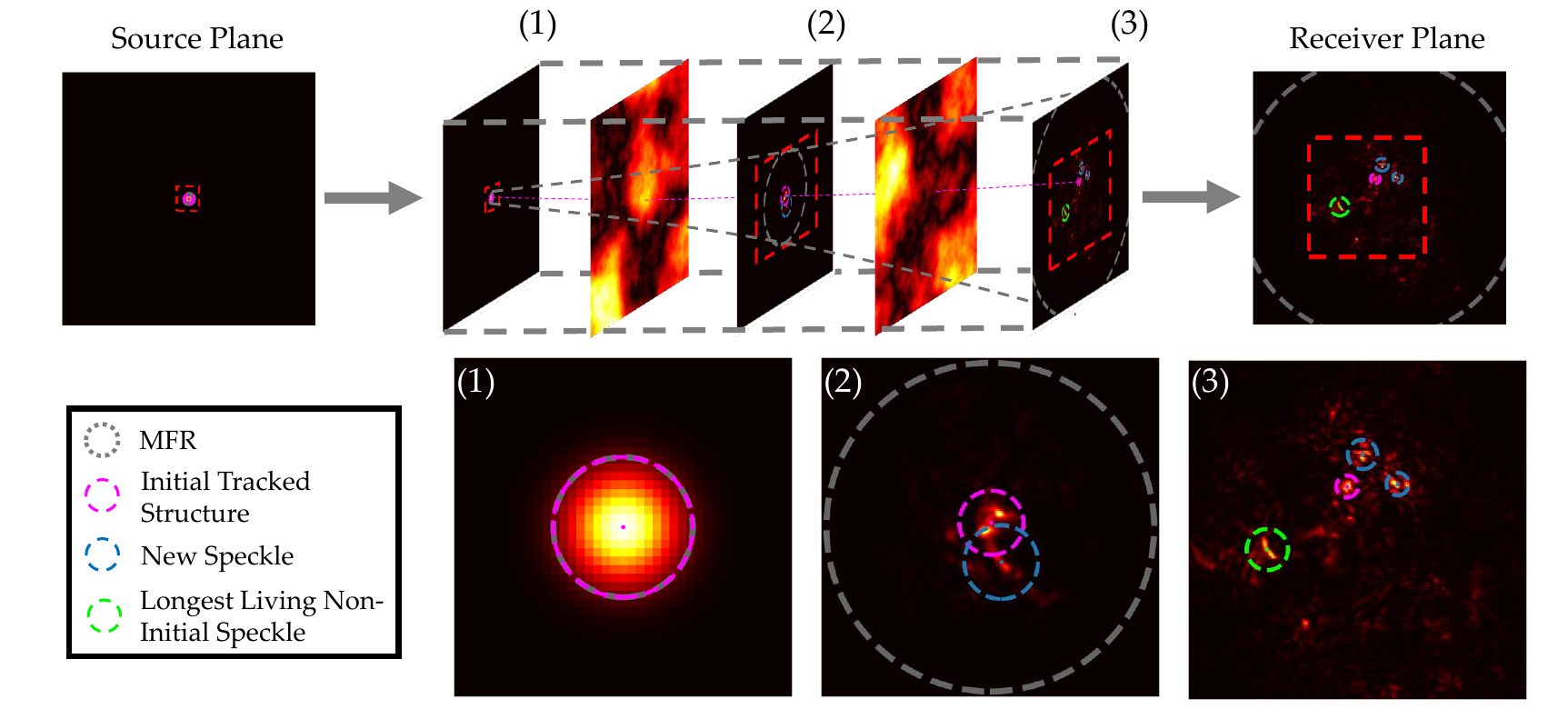}
  \caption{Speckle simulation and tracking. Top row: spatial beam profiles from the source to the receiver plane with representative turbulence profiles. The mode field radius $\tilde{r}(z)$ (gray), source plane tracked structure (magenta), longest lived non-initial speckle (green), and other detected speckles (blue) are tracked based on their nearest neighbor conditions. Bottom row: magnified views (red box) of observation points (1-3). Detected structures have a local peak intensity greater than $\eta I_{\rm max}$.}
  \label{fig:Tracking}
\end{figure*}

\section{Numerical model and analysis methods}

Electromagnetic wave propagation through a turbulent medium is modeled using the inhomogeneous paraxial wave equation \cite{kolokolov2020statistical}
\begin{equation}
  \nabla_\perp^2 U + 2ik\frac{\partial U}{\partial z} + 2k^2 \delta n(x,y,z)\,U = 0,
  \label{phe}
\end{equation}
where $k = 2\pi n_0/\lambda$, $n_0$ is the background refractive index, $\nabla_\perp^2 = \partial^2/\partial x^2 + \partial^2/\partial y^2$, $\delta n(x,y,z)$ represents stochastic refractive-index fluctuations, and $U(x,y,z)$ is the complex-valued electric-field amplitude. Equation \ref{phe} is solved numerically using split-step propagation with the Fresnel transfer function \cite{voelz2011computational}.

Refractive-index fluctuations are generated using the modified von Kármán power spectrum
\begin{equation}
  \Phi_n(\kappa)
  = 0.033\,C_n^2
    \bigl(\kappa^2 + k_0^2\bigr)^{-11/6}
    \exp\!\left(-\frac{\kappa^2}{k_m^2}\right),
    \label{MVK}
\end{equation}
where $\kappa$ is the transverse spatial frequency, $C_n^2$ is the refractive-index structure parameter, $k_0 = 2\pi/L_0$ corresponds to the outer scale $L_0$, and $k_m = 5.92/\ell_0$ corresponds to the inner scale $\ell_0$. Here, $L_0$ and $\ell_0$ are fixed to demonstrate the tracking algorithm.

\subsection{Tracked observables}

To enable object-level analysis, we implement an intensity-based speckle detection and tracking algorithm that identifies and follows individual intensity substructures during propagation. For each propagation realization, the following observables are extracted:
\begin{enumerate}[label=(\roman*)]
  \item $\text{j}^{\text{th}}$ speckle radius $\tilde r_j(z)$,
  \item speckle persistence $\Delta z_j$,
  \item global beam mode field radius $\tilde r(z)$,
  \item number of speckles $N_s$.
\end{enumerate}
Together, these observables characterize both the evolution of individual intensity substructures and the overall transverse spreading of the beam. This represents a subset of quantities accessible within this object-level framework. The results therefore demonstrate the scope of the algorithmic framework and permit extension to additional feature-based observables.

\subsection{Speckle detection}

Here, individual substructures are defined by a finite detection threshold $\eta$. Speckles represent resolvable interconnected collections of pixels whose associated intensity value exceeds a fraction of the source-plane peak intensity $\eta I_{\text{max}}$. $\eta$  thus serves as a minimum contrast criterion for resolvable features. All results presented here, unless specified otherwise, correspond to a fixed threshold of $\eta = 0.025$.

Speckle centers are determined by calculating the centroid of each individual collection of interconnected pixels using
\begin{equation}
\mathbf r_{c,j}(z;\eta)=
\frac{\displaystyle \sum_{(m,n)\in\Omega_j(z;\eta)} 
\mathbf r_{m,n}\, I_{m,n}(z)}
{\displaystyle \sum_{(m,n)\in\Omega_j(z;\eta)} 
I_{m,n}(z)},
\qquad 
\mathbf r_{m,n}=(x_m,y_n).
\end{equation}
$(m,n)$ are pixel indices that are within the set of pixels for the $j^{th}$ interconnected structure while $c$ denotes the observable as being related to the centroid. The domain over which these pixels are summed is defined by 
\begin{equation}
\Omega_j(z;\eta)=
\left\{(m,n)\ \middle|\ 
I_{m,n}(z)>\eta I_{\max}(0)
\right\}
\end{equation}

While $\eta$ is imposed in the algorithmic framework, an analogous detection threshold arises naturally in physical measurement systems through the finite sensitivity and dynamic range of the receiver. In an experimental setting, the effective value of $\eta$ is set to remove background fluctuations associated with the experimental noise floor linked to detector sensitivity. The introduction of $\eta$ in the numerical model thus reflects the practical limitations of feature detectability in real optical systems, ensuring that the tracked substructures correspond to physically resolvable intensity concentrations rather than background fluctuations.

\subsection{Interplanar association and speckle persistence}

Interplanar speckle association is performed between successive propagation planes using nearest-neighbor matching of centroid positions. Let $\mathbf r_{c,j}(z_k)$ denote the centroid of the $j^{\mathrm{th}}$ speckle detected at plane $z_k$ for a fixed detection threshold $\eta$. For each active speckle at $z_k$, candidate centroids at the subsequent plane $z_{k+1}$ are evaluated using Euclidean separation
\begin{equation}
d_{ij}(z_k)=\left\|\mathbf r_{c,i}(z_{k+1})-\mathbf r_{c,j}(z_k)\right\|.
\end{equation}
Association is accepted only when
\begin{equation}
\min_i d_{ij}(z_k)\le r_a,
\end{equation}
where $r_a=15$ pixels is a fixed association radius. This value is chosen as a conservative compromise: large enough to accommodate interplanar centroid displacements of a continuing speckle, yet small enough to suppress false reassociation between distinct intensity substructures. Small perturbations about this value appear to have little effect on the qualitative trends reported here, whereas larger changes on the order of $\pm 5\, \text{pixels}$ produce more instances of premature track termination and false reassociation.

If no centroid satisfies the association criterion, or if the corresponding intensity structure at $z_{k+1}$ no longer satisfies the detection threshold, the speckle track is terminated. Conversely, centroids at $z_{k+1}$ that are not associated with any preexisting track initiate new speckle tracks, provided they do not lie within $r_a$ of an already active centroid. This condition suppresses redundant track formation in regions of elevated local intensity concentration and enforces spatial continuity of the tracked evolution.

The persistence of the $j^{\mathrm{th}}$ speckle is defined as
\begin{equation}
\Delta z_j = z_{\mathrm{death},j} - z_{\mathrm{creation},j},
\end{equation}
where $z_{\mathrm{creation},j}$ and $z_{\mathrm{death},j}$ denote, respectively, the first and last propagation planes at which that speckle is successfully identified and associated. Accordingly, $\Delta z_j$ should be interpreted as a tracking-conditioned propagation distance rather than as an invariant physical lifetime of an underlying field structure. Together, the threshold $\eta$ and association radius $r_a$ define the detection and association criteria under which the object-level observables reported in this work are measured.

\subsection{Tracked speckle radius and mode field radius}

Classical speckle theory characterizes speckle size using second-order spatial statistics, such as the intensity autocorrelation function or mutual coherence function \cite{goodman1975statistical}. Here, we instead define speckle width from the detected intensity substructures. These definitions are not identical, but both quantify a characteristic transverse scale of the speckle field. Atmospheric and phase-screen models predict that this transverse scale decreases as turbulence strength or random phase perturbation strength increases \cite{korotkova2002speckle,korotkova2002speckless}. In related rough-target lidar models, the number of speckle cells is estimated by dividing the collection area by the speckle correlation area \cite{korotkova2002speckle,korotkova2004lidar}. Therefore, although our thresholded connected-component count is not equivalent to a coherence-cell count, the expected qualitative trend is the same. Stronger turbulence produces smaller characteristic speckle scales and more resolvable speckle structures.

For each tracked speckle indexed by $j$, the radius $\tilde r_j(z)$ is defined as the mean radial distance from the speckle centroid to the discrete $1/e^2$ intensity contour at propagation distance $z$. Let $\mathbf r_{c,j}(z)$ denote the centroid of the $j^{\mathrm{th}}$ tracked speckle, and let $\mathbf r_{j,n}(z)$, $n=1,\dots,M_j(z)$, denote the $M_j(z)$ sampled points along the corresponding contour defined by the intensity level $I(\mathbf r_{c,j}(z),z)e^{-2}$. The speckle radius is then given by
\begin{equation}
\tilde r_j(z)
=
\frac{1}{M_j(z)}
\sum_{n=1}^{M_j(z)}
\left\|
\mathbf r_{j,n}(z)-\mathbf r_{c,j}(z)
\right\|.
\end{equation}

To contextualize speckle-level behavior within the evolution of the full optical field, we also compute a global beam radius that characterizes the transverse extent of the entire field. This quantity is taken to be the mode field radius (MFR), defined from the second moment of the intensity distribution. To isolate envelope broadening from beam wander, the intensity distribution at each propagation plane is first translated so that its beam centroid coincides with the optical axis. The MFR is then evaluated as
\begin{equation}
  \tilde r(z)
  = \frac{\sqrt{2\iint (x^2 + y^2)\,|U(x,y,z)|^2\,dx\,dy}}
         {\sqrt{\iint |U(x,y,z)|^2\,dx\,dy}}.
\end{equation}
The MFR measures the overall transverse redistribution of optical power during propagation and captures the global expansion of the beam envelope under the combined influence of diffraction and turbulence.

Unlike the tracked speckle radius, the MFR does not resolve internal intensity structure within the beam. As turbulence fragments the field into smaller intensity substructures, those features remain embedded within the broader envelope characterized by $\tilde r(z)$. The MFR therefore provides a global reference scale against which the evolution of tracked speckle radii and trajectories may be interpreted. 

A representative demonstration of the tracking algorithm is shown in Fig.~\ref{fig:Tracking}, together with the corresponding global MFR. \added{Source code for the speckle tracking algorithm and figure-generation is provided in supplement~1}. In this example, a Gaussian beam propagates through strong turbulence, where each phase screen is generated as an independent Gaussian random realization with statistics prescribed by the modified von Kármán power spectral density of Eq.~\ref{MVK}. At the source plane, the $1/e^2$ radius of the initially identified structure coincides with the beam MFR. As propagation proceeds, the beam fragments into smaller localized substructures whose sizes no longer reflect the global MFR and whose centroids evolve independently across the transverse plane.

\section{Results and Discussion}

\subsection{Global beam radius as a reference metric}

\begin{figure}[ht!]
  \centering
  \includegraphics[width=\linewidth]{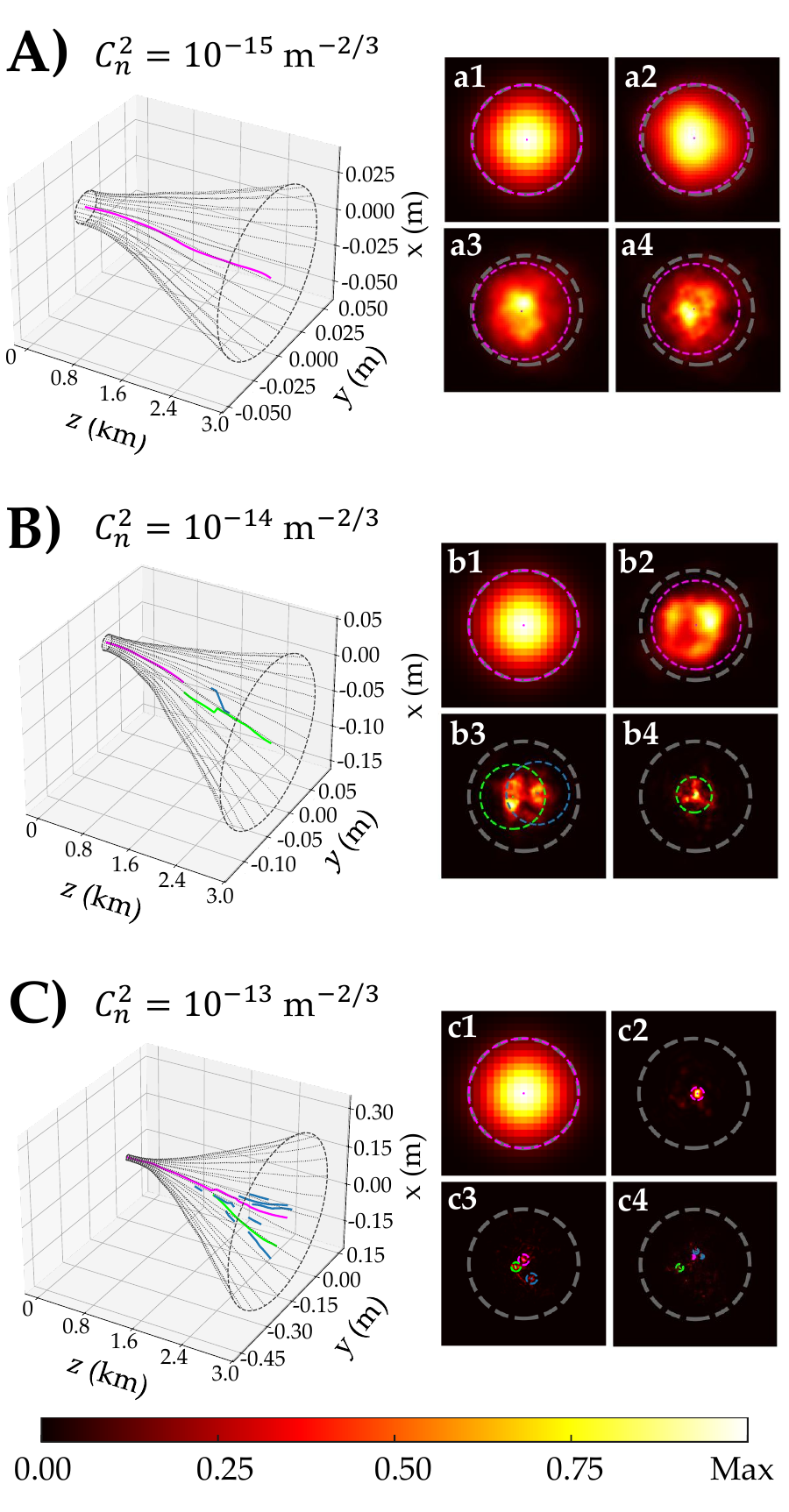}
  \caption{Propagation of Gaussian profiles under (A) weak turbulence, (B) moderately-strong turbulence, and (C) strong turbulence. The transverse profiles (a1–a4, b1–b4, c1–c4) show the evolution of the global mode field radius $\tilde r(z)$ (gray) and embedded speckles (magenta, lime, blue) at $z = 0, 975, 1950, 3000\ \mathrm{m}$ under the conditions of $w_0 = 1\ \mathrm{cm}$ and $\ell_0 = 5\ \mathrm{mm}$. The color bar shows the intensity scale normalized for each transverse profile (a1--a4, b1--b4, c1--c4).}
  \label{fig:modefieldradius}
\end{figure}

Figure~\ref{fig:modefieldradius} shows the evolution of the global mode field radius $\tilde r(z)$ for increasing turbulence strength. In all cases, the MFR increases monotonically with propagation distance, indicating continued broadening of the overall beam envelope. The transverse profiles (a1--a4, b1--b4, c1--c4) show that, while this global envelope continues to expand, the internal intensity distribution becomes increasingly fragmented as turbulence strength increases. The colored trajectory lines visualize the propagation paths of tracked speckles within a single realization.

The MFR therefore serves as a global reference scale for the broader beam envelope within which tracked speckle trajectories evolve. As turbulence strength increases, fragmentation appears closer to the source plane and the transverse spread of tracked speckles broadens within the expanding beam profile. 

This distinction is important because the MFR is a second-moment measure of the full intensity distribution and therefore primarily reflects beam-scale power redistribution. It does not distinguish between smooth broadening of the field and broadening accompanied by internal fragmentation into multiple compact substructures. As a result, monotonic growth of $\tilde r(z)$ can occur simultaneously with the appearance of smaller tracked speckles embedded within the expanding beam envelope. This motivates the use of object-level observables such as speckle radius, speckle persistence, and number of speckles, to quantify the temporal evolution of individual intensity substructures.

\subsection{Speckle radius and spatial confinement}

Figure~\ref{fig:widths} shows the evolution of the ensemble statistics computed across 500 independent realizations for different turbulence strengths and source-plane spot sizes. The average radius of identified speckles, $\langle\tilde r_j(z)\rangle$ (solid line), is compared to the global beam envelope characterized by $\tilde r(z)$ (dashed line) for beams with spot sizes of $w_0 = 0.010\ \mathrm{m}$ (red) and $w_0 = 0.020\ \mathrm{m}$ (blue), with $\ell_0 = 0.005\ \mathrm{m}$.

In the absence of turbulence, the optical field remains spatially coherent and there is no fragmentation. In this limit, tracking identifies a single connected structure whose width increases monotonically with propagation distance according to the diffraction-driven expansion of a Gaussian beam in vacuum, $w(z)=w_0\sqrt{1+\left(\frac{z}{z_0}\right)^2}$, where $z_0$ is the Rayleigh range \cite{saleh2019fundamentals}. Under weak turbulence, little separation is observed between the characteristic radius of tracked intensity structures and the global beam radius, so that $\langle\tilde r_j(z)\rangle$ remains close to $\tilde r(z)$.

Under stronger turbulent propagation, stochastic redistribution of optical intensity leads to the formation of multiple speckles whose transverse widths deviate from the global beam envelope. As turbulence strength increases, the ensemble-averaged speckle radius departs progressively from the monotonic growth of the global MFR. Under moderately strong turbulence, this deviation appears as a reduced growth rate and the onset of flattening, whereas under strong turbulence the tracked speckle radius becomes increasingly bounded over the simulated propagation interval. In Fig.~\ref{fig:widths}C, where $C_n^2 = 10^{-13}\ \mathrm{m}^{-2/3}$, this bounded behavior is evident for both source-plane spot sizes, even as the global MFR continues to increase. This behavior is only observed at the level of tracked speckle substructures and is not reflected in the global MFR.

To further characterize this transition, we compare the atmospheric transverse coherence length to the global MFR using
\begin{equation}
    \chi(z) = \frac{\rho_0(z)}{\langle\tilde{r}(z)\rangle},
\end{equation}
where $\rho_0 = \left(1.46C_n^2k^2z\right)^{-3/5}$. Here, smaller values of $\chi$ indicate that the beam spans more transverse coherence lengths. In weak turbulence, there is no bounding behavior since there is little speckle formation. For $C_n^2=10^{-14}\ \text{m}^{-2/3}$, the bounded behavior appears to occur at around $z \approx2\ \text{km}$ ($\chi=5.57$) for $w_0 = 0.010\ \text{m}$ and $z\approx 2.5\ \text{km}$ ($\chi=5.41$) for $w_0 = 0.02\ \text{m}$. For $C_n^2 = 10^{-13}\ \text{m}^{-2/3}$ , the bounded trend occurs closer to $z\approx1\ \text{km}$ ($\chi=0.86$) for $w_0 = 0.01\ \text{m}$ and $z\approx 0.25\ \text{km}$ ($\chi = 1.92$) for $w_0 = 0.020\ \text{m}$.

The fluctuations in $\tilde r_j(z)$ are consistent with scintillation-driven redistribution of optical power \cite{andrews1999theory,andrews2005laser,andrews2023laser,stotts2024optical} and with local phase-dependent structure within the speckle field \cite{angelsky2009spatial,cheng2014phase}. When optical power becomes more strongly concentrated within an identified speckle, its associated $1/e^2$ contour contracts and $\tilde r_j(z)$ decreases. When that concentration weakens, the measured speckle radius expands. Accordingly, $\tilde r_j(z)$ reflects local intensity reorganization at the object level rather than the global beam spreading captured by $\tilde r(z)$.

\subsection{Speckle persistence statistics}

\begin{figure}[ht!]
  \centering
  \includegraphics[width=\linewidth]{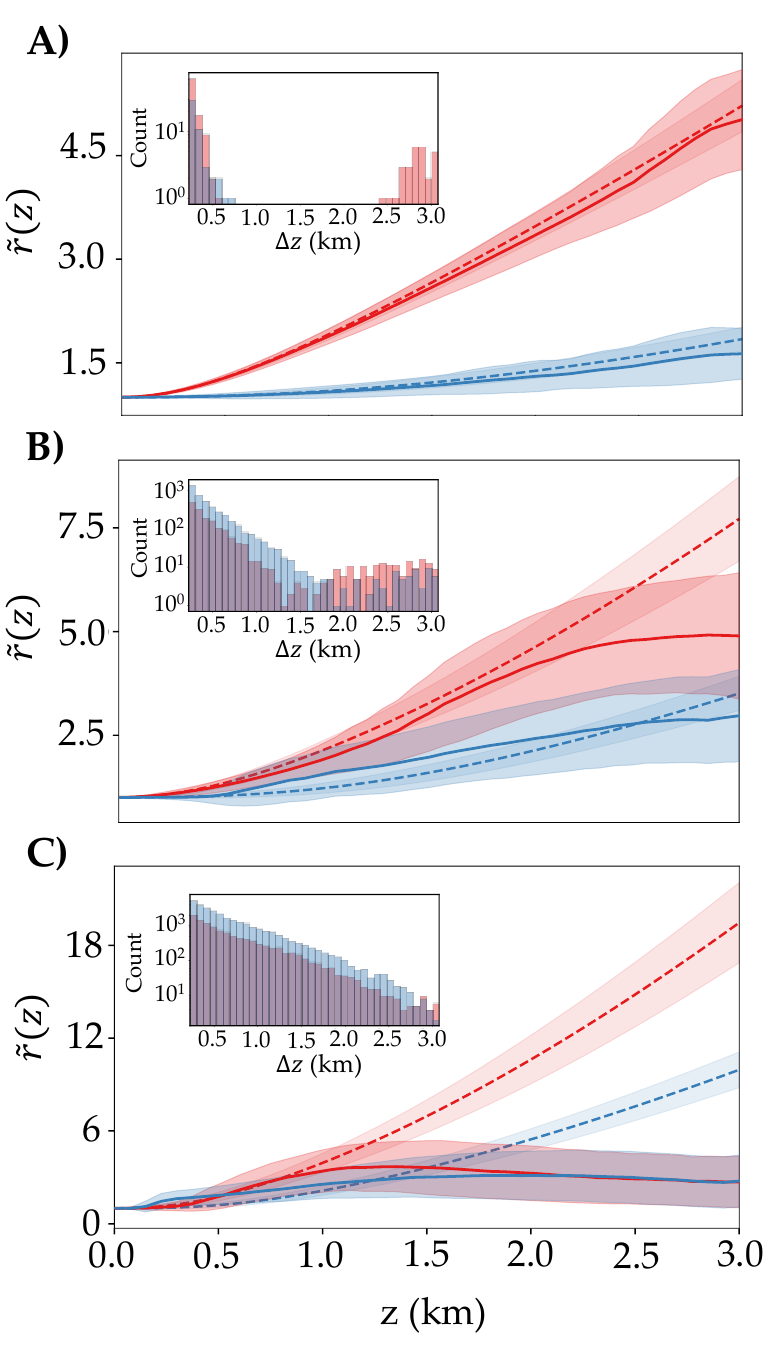}
  \caption{Evolution of the ensemble-averaged spot-normalized MFR $\tilde{r}(z)/w_0$ of the global beam (dashed line), the ensemble-averaged normalized $1/e^2$ radius $\langle\tilde{r}_j(z)\rangle/w_0$ of the tracked speckles (solid line), and standard deviation $\pm1\sigma$ (shaded regions) under (A) weak, (B) moderately-strong, and (C) strong turbulence. Beams with widths $w_0 = 0.010\ \text{m}$ (red) and $w_0 = 0.020\ \text{m}$ (blue) are shown. Inset histograms show the distribution of speckle persistence $\Delta z_j$.}
  \label{fig:widths}
\end{figure}

The inset histograms in Fig.~\ref{fig:widths} show the distribution of speckle persistence distances $\Delta z$ for beams with initial spot sizes $w_0 = 0.010\ \mathrm{m}$ (red) and $w_0 = 0.020\ \mathrm{m}$ (blue) over a total propagation length of $L = 3.0\ \mathrm{km}$. Only structures that both appear and disappear within the propagation interval are included in the histograms. Tracks that persist through the full propagation length are excluded since their termination distance cannot be determined.

Under weak turbulence [Fig.~\ref{fig:widths}A, $C_n^2 = 10^{-15}\ \mathrm{m}^{-2/3}$], the field is largely dominated by the source-plane structure in each realization, with relatively few additional detections forming during propagation. The persistence distribution is therefore bimodal, reflecting that most tracked structures that do not survive the full path are either short lived fragments or long lived structures that disappeared just before the end of the propagation path. In this regime turbulence introduces only minor fragmentation of the optical field.

As turbulence strength increases [Fig.~\ref{fig:widths}B, $C_n^2 = 10^{-14}\ \mathrm{m}^{-2/3}$], additional speckles appear throughout the beam and persistence distances become more broadly distributed. The histograms transition to a mixed population spanning a wide range of lifetimes. While a substantial number of tracks still persist over kilometer-scale distances, many structures now form and disappear within intermediate portions of the propagation path, reflecting the increasing influence of turbulence-induced intensity fluctuations.

In the strong turbulence regime [Fig.~\ref{fig:widths}C, $C_n^2 = 10^{-13}\ \mathrm{m}^{-2/3}$], the persistence distributions shift decisively toward shorter lifetimes. The field becomes populated by a large number of speckles that appear and disappear over relatively short propagation intervals. Long-lived structures remain present but constitute a much smaller fraction of the detected population.

These results indicate that increasing turbulence strength drives a transition from a regime dominated by long-lived structures to one characterized by frequent formation and disappearance of speckles throughout the beam. The persistence statistics therefore provide a complementary description of the fragmentation dynamics observed in the transverse width evolution of Fig.~\ref{fig:widths}.

\subsection{Detection-threshold scaling of speckle populations}

The detection threshold $\eta$ defines the minimum intensity required for a substructure to be identified as a speckle. While $\eta$ enters the algorithm as a numerical parameter, it also corresponds physically to the finite dynamic range and sensitivity of an optical receiver, which limits the minimum detectable contrast level. Varying $\eta$ therefore probes how the population of resolvable intensity structures depends on the detectability criterion.

\begin{figure}[ht!]
    \centering
    \includegraphics[width=1\linewidth]{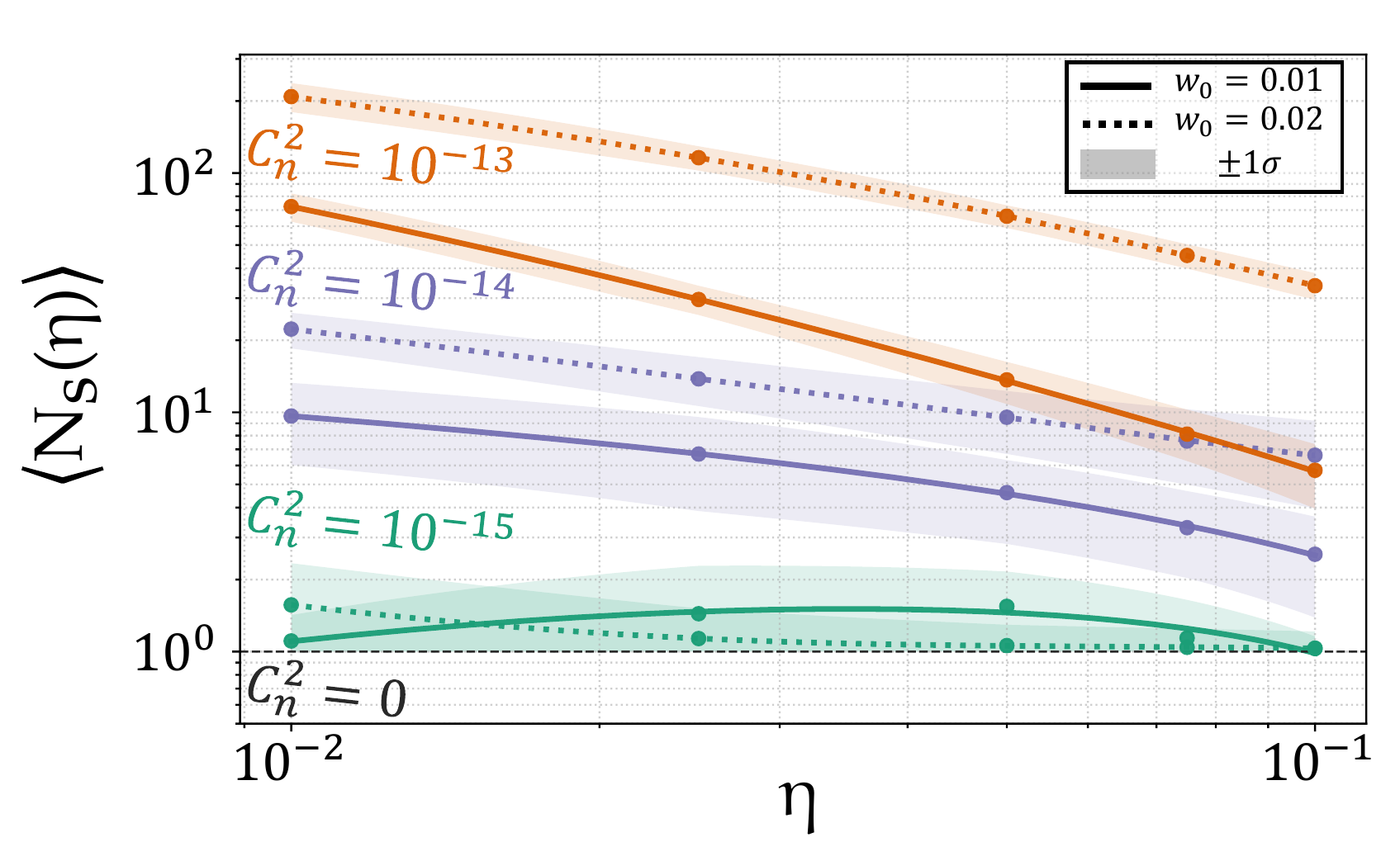}
    \caption{Ensemble-averaged number of detected speckles $\langle N_s(\eta)\rangle$ as a function of detection threshold $\eta$ for different turbulence strengths. Solid lines correspond to $w_0=0.010\,\mathrm{m}$ and dotted lines to $w_0=0.020\,\mathrm{m}$. Points show ensemble means over 500 realizations and shaded regions denote $\pm1\sigma$. The dashed line indicates the vacuum case ($C_n^2=0$), for which only a single connected structure is detected. }
    \label{fig:NvsEta}
\end{figure}

To examine this dependence, we compute the ensemble-averaged number of detected speckles $\langle N_s(\eta) \rangle$, where $N_s$ denotes the total number of speckles detected in a single realization over the full propagation interval. The ensemble average is taken over 500 independent turbulence realizations for each set of parameters. Figure~\ref{fig:NvsEta} shows the dependence of $\langle N_s(\eta) \rangle$ on the detection threshold for multiple turbulence strengths and beam widths.

In the absence of turbulence ($C_n^2 = 0$, black dashed line), the optical field remains spatially coherent and only a single connected intensity structure is detected, giving $\langle N_s \rangle = 1$ independent of $\eta$. Under weak turbulence ($C_n^2 = 10^{-15}\,\mathrm{m}^{-2/3}$), the detected population remains close to unity across the explored threshold range, consistent with limited fragmentation and the persistence of a dominant source-plane structure. However, the weak-turbulence curves also exhibit a slight non-monotonic dependence on $\eta$. This behavior arises from the thresholded connected-component definition of a speckle: at lower thresholds, neighboring lobes may remain connected through weak intensity bridges and be counted as a single structure, whereas at intermediate thresholds those bridges can be removed, causing the same region to split into multiple detected components. As the threshold is increased further, weaker components may then fall below detectability entirely, reducing the count again.

For intermediate turbulence ($C_n^2 = 10^{-14}\,\mathrm{m}^{-2/3}$), the detected population grows more noticeably as $\eta$ decreases, indicating the formation of additional lower-intensity structures within the beam envelope. In the strong turbulence regime ($C_n^2 = 10^{-13}\,\mathrm{m}^{-2/3}$), $\langle N_s(\eta) \rangle$ shows a much steeper dependence on $\eta$, consistent with the appearance of many additional detectable structures as the threshold is lowered. At each turbulence strength, the larger source-plane beam ($w_0=0.020\,\mathrm{m}$) yields a systematically larger detected population than the smaller beam ($w_0=0.010\,\mathrm{m}$) over the threshold range shown.

These results show that the threshold dependence of the detected speckle population is shaped both by the underlying intensity fluctuations in the turbulent field and by the connectivity criterion imposed by the detection threshold. Accordingly, $\langle N_s(\eta)\rangle$ provides a useful threshold-conditioned observable for comparing how fragmentation develops across different propagation conditions.

\section{Conclusion}

We introduce an object-level framework for analyzing turbulent speckle fields by detecting and tracking connected intensity substructures along the propagation path. This approach extends the description of turbulent propagation beyond purely global ensemble-averaged beam metrics by constructing observables defined on individual tracked speckles. Within this framework, we quantify (i) transverse speckle radii from $1/e^2$ contours about each tracked centroid, (ii) persistence distances $\Delta z_j$ from the first and last successful detections of a tracked speckle, and (iii) trajectories of identified structures within the globally expanding beam envelope characterized by the mode field radius $\tilde r(z)$.

Across all turbulence strengths considered here, the tracked speckle radii exhibit distinctly different behavior: under weak turbulence they monotonically grow with the MFR, while under stronger turbulence they become increasingly bounded over the simulated propagation interval. The persistence statistics similarly evolve with turbulence strength, shifting from populations containing a larger fraction of long-lived detections in weak turbulence to populations dominated by shorter-lived detections in strong turbulence. These results demonstrate a clear distinction between global beam expansion and the evolution of internal tracked substructures.

We also show that the ensemble-averaged number of detected speckles, $\langle N_s(\eta)\rangle$, depends strongly on the detection threshold $\eta$, turbulence strength $C_n^2$, and source plane beam width $w_0$. In weak turbulence, $\langle N_s(\eta)\rangle$ remains close to unity, with only slight non-monotonicity arising from threshold-dependent changes to the connected-component topology. In stronger turbulence, the detected speckle population grows much more rapidly as the threshold is lowered and the source plane beam width is increased. This behavior shows that the dependence of the detected speckle counts on $\eta$ and $w_0$ carries meaningful information about how turbulence redistributes optical intensity across the beam-sampled turbulent field.

More broadly, these results show that object-level observables such as tracked speckle radius, persistence distance, and population counts provide a useful complement to conventional beam-scale statistics in turbulent propagation. The present framework therefore provides a practical basis for studying how internal intensity structure forms, evolves, and disappears within distorted optical fields, and it opens the door to additional object-level observables beyond those considered here.

\section*{Appendix I: Summary of Observables}

For clarity, we summarize the global and object-level observables used throughout this work in Table~\ref{tab:metrics_summary}. These quantities are evaluated either directly from the propagated optical field or from tracked intensity substructures identified through the detection procedure described in Sec.~2. Global metrics characterize the transverse redistribution of optical energy within the full beam envelope, while object-level observables quantify the geometric and persistence properties of resolvable speckle structures throughout propagation.

\begin{table*}[t]
\centering
\renewcommand{\arraystretch}{1.75}
\setlength{\tabcolsep}{5pt}
\small
\begin{tabular}{>{\raggedright\arraybackslash}p{0.20\textwidth}
                >{\raggedright\arraybackslash}p{0.36\textwidth}
                >{\raggedright\arraybackslash}p{0.40\textwidth}}
\hline
\textbf{Metric} & \textbf{Definition} & \textbf{Physical Interpretation} \\
\hline
Intensity &
$I(x,y,z)=|U(x,y,z)|^2$ &
Optical power distribution in the transverse plane. \\[6pt]

Global MFR &
$\displaystyle
\tilde r(z)=\sqrt{\frac{2\iint (x^2+y^2)|U(x,y,z)|^2\,dx\,dy}{\iint |U(x,y,z)|^2\,dx\,dy}}$ &
Second-moment beam radius describing global transverse energy redistribution. \\[10pt]

Speckle radius &
$\displaystyle
\tilde r_j(z)=\frac{1}{M_j(z)}
\sum_{n=1}^{M_j(z)}
\|\mathbf r_{j,n}(z)-\mathbf r_{c,j}(z)\|$ &
Average transverse extent of the detected $1/e^2$ contour of speckle $j$. \\[10pt]

Speckle persistence &
$\Delta z_j = z_{\mathrm{death},j} - z_{\mathrm{creation},j}$ &
Tracking-conditioned propagation distance over which speckle $j$ remains detectable. \\[8pt]

Detection threshold &
$I_{\mathrm{th}}=\eta I_{\max}$ &
Minimum resolvable local intensity relative to the source-plane peak. \\
\hline
\end{tabular}
\caption{Summary of global and object-level observables used throughout this work.}
\label{tab:metrics_summary}
\end{table*}

\begin{figure}
    \centering
    \includegraphics[width=\linewidth]{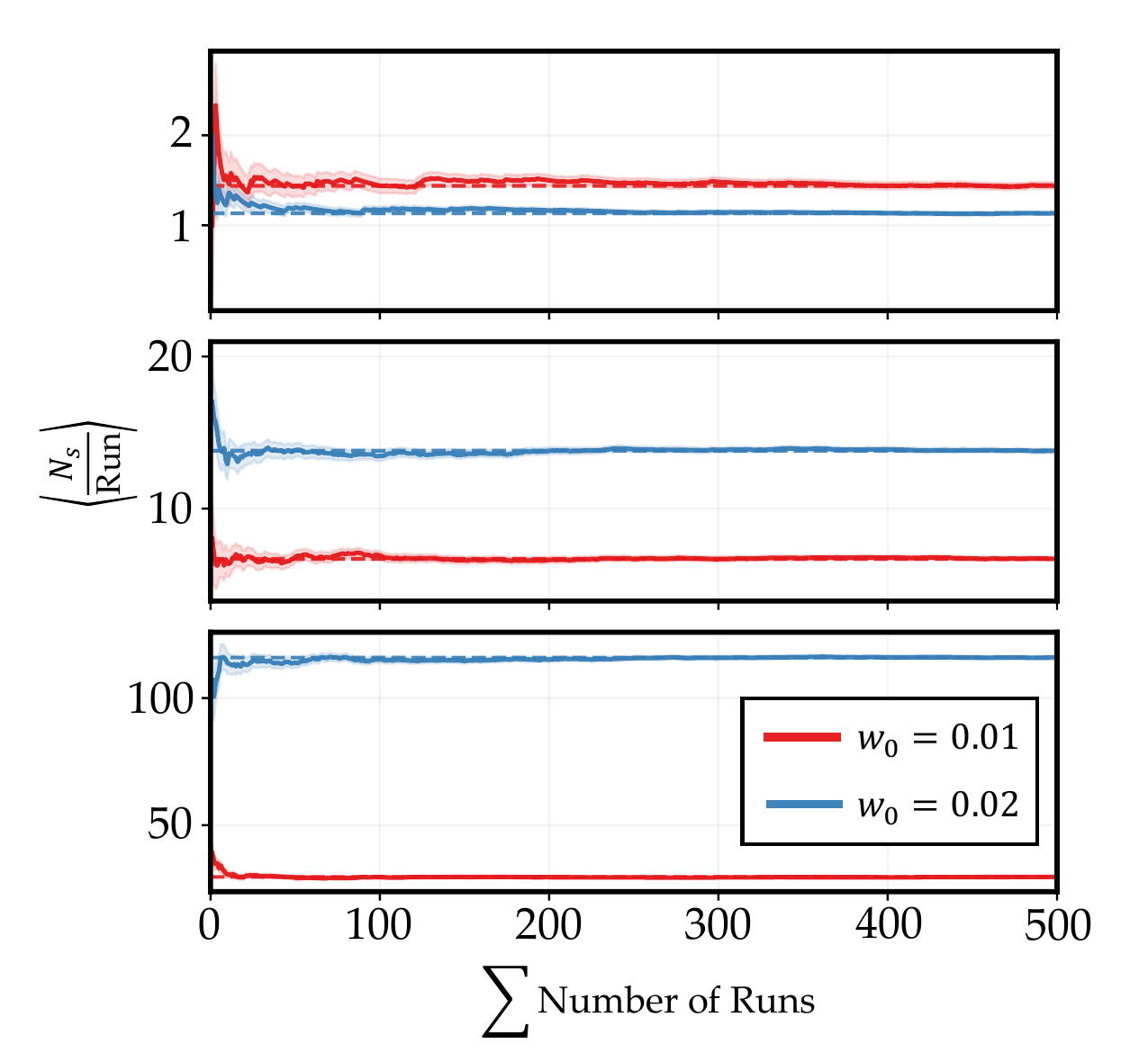}
    \caption{Cumulative mean of the total number of detected speckles per realization as a function of the number of independent turbulence realizations for beams with $w_0 = 0.010~\mathrm{m}$ (red) and $w_0 = 0.020~\mathrm{m}$ (blue) under weak (top), moderate (middle), and strong (bottom) turbulence. Shaded regions denote $\pm 1$ standard error of the mean.}
    \label{fig:N_per_run}
\end{figure}

\section*{Appendix II: Convergence of speckle statistics}

Reliable estimation of object-level statistics requires that the detected speckle population be sampled over a sufficient number of independent turbulence realizations. In particular, observables such as persistence, transverse radius, and number of speckles depend on the number of resolvable intensity substructures identified in each realization. Insufficient sampling may therefore bias the inferred speckle population and the associated lifetime statistics.

To verify that the reported results are obtained within a statistically stable regime, we examine the cumulative mean number of detected speckles per realization as a function of the total number of runs included in the ensemble. Figure~\ref{fig:N_per_run} shows the convergence behavior of this quantity for each turbulence strength and beam width considered in this study. In all cases, the cumulative mean approaches a stable value as additional realizations are incorporated, indicating that the detected speckle population has been sampled sufficiently for the associated statistics to be representative of the underlying turbulent field. All subsequent results are therefore reported using ensembles operating within this converged regime.

The statistical uncertainty reported in Fig. \ref{fig:N_per_run} is evaluated from the run-to-run variability across independent turbulence realizations. Let $N_i$ denote the number of detected speckles in the $i^{\mathrm{th}}$ realization, and let $M$ denote the total number of realizations. The mean number of speckles per realization is then
\begin{equation}
\bar{N} = \frac{1}{M} \sum_{i=1}^{M} N_i,
\end{equation}
with standard error of the mean given by
\begin{equation}
SE_{\bar{N}} = \frac{s}{\sqrt{M}},
\end{equation}
where $s$ is the sample standard deviation of $\{N_i\}$,
\begin{equation}
s = \sqrt{\frac{1}{M-1} \sum_{i=1}^{M} \left(N_i - \bar{N}\right)^2}.
\end{equation}

\section*{Acknowledgments}
The authors gratefully acknowledge funding from the United States Air Force Office of Scientific Research (A9550-24-1-0027 P00002).

\section*{Disclosures}
The authors declare no conflicts of interest.

\section*{Data availability}
Datasets are available upon reasonable request. 

\section*{Supplemental Documentation}
See supplement 1 for data and code. 

\bibliographystyle{unsrtnat}
\bibliography{Refs}

@book{saleh2019fundamentals,
  title={Fundamentals of photonics, 2 volume set},
  author={Saleh, Bahaa EA and Teich, Malvin Carl},
  year={2019},
  publisher={john Wiley \& sons}
}

@article{long2023underwater,
  title={Underwater forward-looking sonar images target detection via speckle reduction and scene prior},
  author={Long, Hui and Shen, Liquan and Wang, Zhengyong and Chen, Jinbo},
  journal={IEEE Transactions on Geoscience and Remote Sensing},
  volume={61},
  pages={1--13},
  year={2023},
  publisher={IEEE}
}

@article{goodman1976some,
  title={Some fundamental properties of speckle},
  author={Goodman, Joseph W},
  journal={JOSA},
  volume={66},
  number={11},
  pages={1145--1150},
  year={1976},
  publisher={Optica Publishing Group}
}

@book{goodman2007speckle,
  title={Speckle phenomena in optics: theory and applications},
  author={Goodman, Joseph W},
  year={2007},
  publisher={Roberts and Company Publishers}
}

@incollection{goodman1975statistical,
  title={Statistical properties of laser speckle patterns},
  author={Goodman, Joseph W},
  booktitle={Laser speckle and related phenomena},
  pages={9--75},
  year={1975},
  publisher={Springer}
}

@inproceedings{korotkova2004lidar,
  title={LIDAR model for a rough-surface target: Method of partial coherence},
  author={Korotkova, Olga and Andrews, Larry C and Phillips, Ronald L},
  booktitle={Optics in Atmospheric Propagation and Adaptive Systems VI},
  volume={5237},
  pages={49--60},
  year={2004},
  organization={SPIE}
}

@inproceedings{korotkova2002speckless,
  title={Speckle propagation through atmospheric turbulence: effects of a random phase screen at the source},
  author={Korotkova, Olga and Andrews, Larry C and Phillips, Ronald L},
  booktitle={Free-Space Laser Communication and Laser Imaging II},
  volume={4821},
  pages={98--109},
  year={2002},
  organization={SPIE}
}

@article{lee1976statistics,
  title={Statistics of speckle propagation through the turbulent atmosphere},
  author={Lee, Myung Hun and Holmes, J Fred and Kerr, J Richard},
  journal={Journal of the Optical Society of America},
  volume={66},
  number={11},
  pages={1164--1172},
  year={1976},
  publisher={Optical Society of America}
}

@article{guyot2004spatial,
  title={Spatial speckle characterization by Brownian motion analysis},
  author={Guyot, Steve and P{\'e}ron, Marie-C{\'e}cile and Del{\'e}chelle, Eric},
  journal={Physical Review E--Statistical, Nonlinear, and Soft Matter Physics},
  volume={70},
  number={4},
  pages={046618},
  year={2004},
  publisher={APS}
}

@article{ozcan2007speckle,
  title={Speckle reduction in optical coherence tomography images using digital filtering},
  author={Ozcan, Aydogan and Bilenca, Alberto and Desjardins, Adrien E and Bouma, Brett E and Tearney, Guillermo J},
  journal={Journal of the Optical Society of America A},
  volume={24},
  number={7},
  pages={1901--1910},
  year={2007},
  publisher={Optical Society of America}
}

@article{kayahan2010measurement,
  title={Measurement of surface roughness of metals using binary speckle image analysis},
  author={Kayahan, Ersin and Oktem, Hasan and Hacizade, Fikret and Nasibov, Humbat and Gundogdu, Ozcan},
  journal={Tribology International},
  volume={43},
  number={1-2},
  pages={307--311},
  year={2010},
  publisher={Elsevier}
}

@article{zhang2007speckle,
  title={Speckle evolution of diffusive and localized waves},
  author={Zhang, Sheng and Hu, Bing and Sebbah, Patrick and Genack, Azriel Z},
  journal={Physical review letters},
  volume={99},
  number={6},
  pages={063902},
  year={2007},
  publisher={APS}
}

@inproceedings{macintosh2005speckle,
  title={Speckle lifetimes in high-contrast adaptive optics},
  author={Macintosh, Bruce and Poyneer, Lisa and Sivaramakrishnan, Anand and Marois, Christian},
  booktitle={Astronomical Adaptive Optics Systems and Applications II},
  volume={5903},
  pages={170--177},
  year={2005},
  organization={SPIE}
}

@article{van2020improved,
  title={Improved adaptive-optics performance using polychromatic speckle mitigation},
  author={Van Zandt, Noah R and Spencer, Mark F},
  journal={Applied Optics},
  volume={59},
  number={4},
  pages={1071--1081},
  year={2020},
  publisher={Optical Society of America}
}

@book{andrews2023laser,
  title={Laser beam propagation in random media: new and advanced topics},
  author={Andrews, Larry C and Beason, Melissa K},
  publisher = {SPIE},
  year={2023}
}

@article{migliaccio2019sar,
  title={SAR speckle dependence on ocean surface wind field},
  author={Migliaccio, Maurizio and Huang, Lanqing and Buono, Andrea},
  journal={IEEE Transactions on Geoscience and Remote Sensing},
  volume={57},
  number={8},
  pages={5447--5455},
  year={2019},
  publisher={IEEE}
}

@inproceedings{migliaccio2024link,
  title={On the link between sea surface roughness and speckle distribution under different wind regimes},
  author={Migliaccio, M and Abbasi, F and Zahribanhesari, M and Inserra, G and Verlanti, A and Grieco, G},
  booktitle={IGARSS 2024-2024 IEEE International Geoscience and Remote Sensing Symposium},
  pages={1773--1776},
  year={2024},
  organization={IEEE}
}

@article{andrews1999theory,
  title={Theory of optical scintillation},
  author={Andrews, Larry C and Phillips, Ronald L and Hopen, Cynthia Y and Al-Habash, MA},
  journal={Journal of the Optical Society of America A},
  volume={16},
  number={6},
  pages={1417--1429},
  year={1999},
  publisher={Optical Society of America}
}

@article{stotts2024optical,
  title={Optical communications in turbulence: a tutorial},
  author={Stotts, Larry B and Andrews, Larry C},
  journal={Optical Engineering},
  volume={63},
  number={4},
  pages={041207--041207},
  year={2024},
  publisher={Society of Photo-Optical Instrumentation Engineers}
}

@article{garnier2018imaging,
  title={Imaging through a scattering medium by speckle intensity correlations},
  author={Garnier, Josselin and S{\o}lna, Knut},
  journal={Inverse Problems},
  volume={34},
  number={9},
  pages={094003},
  year={2018},
  publisher={IOP Publishing}
}

@article{garcia2025experimental,
  title={Experimental classification of dynamic speckle regimes: insights from controlled rotational diffuser measurements},
  author={Garcia-Caurel, Enrique and Plyer, Aur{\'e}lien and Colin, Elise and Orlik, Xavier and Ossikovski, Razvigor},
  journal={Journal of the Optical Society of America A},
  volume={42},
  number={10},
  pages={1531--1543},
  year={2025},
  publisher={Optica Publishing Group}
}

@article{gladysz2008detection,
  title={Detection of faint companions through stochastic speckle discrimination},
  author={Gladysz, Szymon and Christou, Julian C},
  journal={The Astrophysical Journal},
  volume={684},
  number={2},
  pages={1486},
  year={2008},
  publisher={IOP Publishing}
}

@inproceedings{korotkova2002speckle,
  title={Speckle propagation through atmospheric turbulence: effects of partial coherence of the target},
  author={Korotkova, Olga and Andrews, Larry C},
  booktitle={Laser Radar Technology and Applications VII},
  volume={4723},
  pages={73--84},
  year={2002},
  organization={SPIE}
}

@inproceedings{korotkova2002speckle2,
  title={Speckle propagation through atmospheric turbulence: effects of a random phase screen at the source},
  author={Korotkova, Olga and Andrews, Larry C and Phillips, Ronald L},
  booktitle={Free-Space Laser Communication and Laser Imaging II},
  volume={4821},
  pages={98--109},
  year={2002},
  organization={SPIE}
}

@inproceedings{beason2019statistical,
  title={Statistical comparison of probability models of intensity fluctuation},
  author={Beason, Melissa and Andrews, Larry and Gladysz, Szymon},
  booktitle={Environmental Effects on Light Propagation and Adaptive Systems II},
  volume={11153},
  pages={104--120},
  year={2019},
  organization={SPIE}
}

@article{kolokolov2020statistical,
  title={Statistical properties of a laser beam propagating in a turbulent medium},
  author={Kolokolov, Igor and Lebedev, Vladimir and Lushnikov, Pavel M},
  journal={Physical Review E},
  volume={101},
  number={4},
  pages={042137},
  year={2020},
  publisher={APS}
}

@article{andrews2005laser,
  title={Laser beam propagation through random media},
  author={Andrews, Larry C and Phillips, Ronald L},
  journal={Laser Beam Propagation Through Random Media: Second Edition},
  year={2005}
}

@article{voelz2011computational,
  title={Computational fourier optics: a MATLAB tutorial},
  author={Voelz, David G},
  journal={(No Title)},
  pages={51},
  year={2011},
  publisher={Spie}
}

@article{liguo2023spatiotemporal,
  title={Spatiotemporal characteristics of dynamic speckle from a 3D target in atmospheric turbulence},
  author={Liguo, Wang and Lei, Gong and Yaqing, Li and Zhiqiang, Yang and Lihong, Yang and Yao, Li},
  journal={Heliyon},
  volume={9},
  number={2},
  year={2023},
  publisher={Elsevier}
}

@article{goldfischer1965autocorrelation,
  title={Autocorrelation function and power spectral density of laser-produced speckle patterns},
  author={Goldfischer, Lester I},
  journal={Journal of the optical society of america},
  volume={55},
  number={3},
  pages={247--253},
  year={1965},
  publisher={Optical Society of America}
}

@article{cheng2014phase,
  title={Phase singularity diffusion},
  author={Cheng, Xiaojun and Lockerman, Yitzchak and Genack, Azriel Z},
  journal={Optics letters},
  volume={39},
  number={11},
  pages={3348--3351},
  year={2014},
  publisher={Optical Society of America}
}

@article{angelsky2009spatial,
  title={Spatial behaviour of singularities in Fractal-and Gaussian speckle fields},
  author={Angelsky, Oleg V and Maksimyak, Alexander P and Maksimyak, Peter P and Hanson, Steen Gr{\"u}ner},
  journal={Open Optics Journal},
  volume={3},
  pages={29--43},
  year={2009},
  publisher={Bentham Open}
}

\end{document}